\documentclass[runningheads,a4paper]{llncs}
\usepackage{xcolor}
\usepackage{times}
\usepackage{amssymb}
\usepackage{amsmath}
\usepackage{graphicx}
\usepackage{url}
\usepackage{booktabs}

\newcommand{\keywords}[1]{\par\addvspace\baselineskip
\noindent\keywordname\enspace\ignorespaces#1}

\urldef{\mailsa}\path|{swayar, ssomit, rrbilgi, srigar}@amazon.com|

\newcommand{\repeatthanks}{\textsuperscript{\thefootnote}} 
\usepackage{array}
\newcolumntype{P}[1]{>{\centering\arraybackslash}p{#1}}

\begin{document}

\title{Improving RNN-T ASR Performance with Date-Time and Location Awareness}

\titlerunning{Context Aware RNN-T ASR}

\author{Swayambhu Nath Ray \thanks{Equal contribution} \and Soumyajit Mitra \repeatthanks \and Raghavendra Bilgi \repeatthanks \and Sri Garimella}

 \authorrunning{Ray, Mitra, Bilgi et al.}
\institute{Alexa Speech, Amazon, India \\
\mailsa\\
}
\index{Ray, Swayambhu Nath}
\index{Mitra, Soumyajit}
\index{Bilgi, Raghavendra}
\index{Garimella, Sri}
\toctitle{} \tocauthor{}

\maketitle

\begin{abstract}
In this paper, we explore the benefits of incorporating context into a Recurrent Neural Network (RNN-T) based Automatic Speech Recognition (ASR) model to improve the speech recognition for virtual assistants. Specifically, we use meta information extracted from the time at which the utterance is spoken and the approximate location information to make ASR context aware. We show that these contextual information, when used individually, improves overall performance by as much as 3.48\% relative to the baseline and when the contexts are combined, the model learns complementary features and the recognition improves by 4.62\%. On specific domains, these contextual signals show improvements as high as 11.5\%, without any significant degradation on others. We ran experiments with models trained on data of sizes 30K hours and 10K hours. We show that the scale of improvement with the 10K hours dataset is much higher than the one obtained with 30K hours dataset. Our results indicate that with limited data to train the ASR model, contextual signals can improve the performance significantly.
\end{abstract}

\section{Introduction}
\label{sec:intro}
Humans often use contextual information to disambiguate a particular utterance and understand incoming speech. The contextual information forms prior knowledge which can be the knowledge about a particular user or world knowledge acquired from many users. In use cases such as voice assistants, there is a lot of prior information about ASR queries. Since we train ASR on data collected from multiple users, which have been said at different contexts, some contextual information is implicitly captured and learned by the model. However, effective use of context may further improve ASR performance. For RNN-T based ASR, there is not much prior art in leveraging contextual information such as state of the device, dialog state, time at which the utterance was spoken, and state or country of origin etc.

In this paper, we focus on providing date-time and geographical information to RNN-T based ASR \cite{chan2016listen,graves2012sequence,graves2006connectionist}. We hypothesize that date-time can be an useful signal for ASR as it carries information about type of utterances, e.g. Christmas related queries will occur frequently in December. Similarly, geographical location may encapsulate user accent, and therefore benefits ASR. We demonstrate the efficacy of explicitly providing contextual information to RNN-T based ASR using up to 30K hours of de-identified queries from smart speakers.

The rest of the paper is organized as follows. We review prior work around the use of context in end-to-end (E2E) ASR in Section \ref{sec:priorwork}. Our context representation techniques and details of the models are outlined in Section \ref{sec:context-representation}. Section \ref{sec:setup} contains the experimental details. Results and discussions are presented in Section \ref{sec:results-and-discussion}. Finally, Section \ref{sec:conclusion} concludes the paper.

\section{Prior Work}
\label{sec:priorwork}

In the literature, contextual information has been successfully used in the language modelling. In \cite{scheiner2016ngramcontext}, location and spoken queries are used for on-the-fly adaptation of the n-gram language model. In neural models, context is often supplied either via embeddings or one-hot vectors. In \cite{mikolov_rnnlm_2021}, RNN language model is adapted based on input contextual information. Where as in \cite{lowrankcontextaware}, context embeddings are used to control a low-rank transformation of the recurrent layer weight matrix. For document classification task, temporal information has been shown to be useful \cite{vashishth2018dating}. Explicitly extracting contextual information also improved the results \cite{ray2018ad3}. In Knowledge graphs, time information is used to learn relation between entities \cite{dasgupta-etal-2018-hyte}. For RNN-T ASR, using intent based semantic signals has been shown to improve performance \cite{ray2021listen}. In \cite{wu2020multistate} contextual meta data such as music playing state and dialog state information has been explored. Since dialog state information is available only at the end of first turn (or utterance), it can be applied to improve recognition of subsequent turns (or utterances). Where as, in our work presented here, we explore using context that is applicable to all utterances.

\begin{figure}[h]
\centering
  \includegraphics[scale=0.4,width=0.4\linewidth]{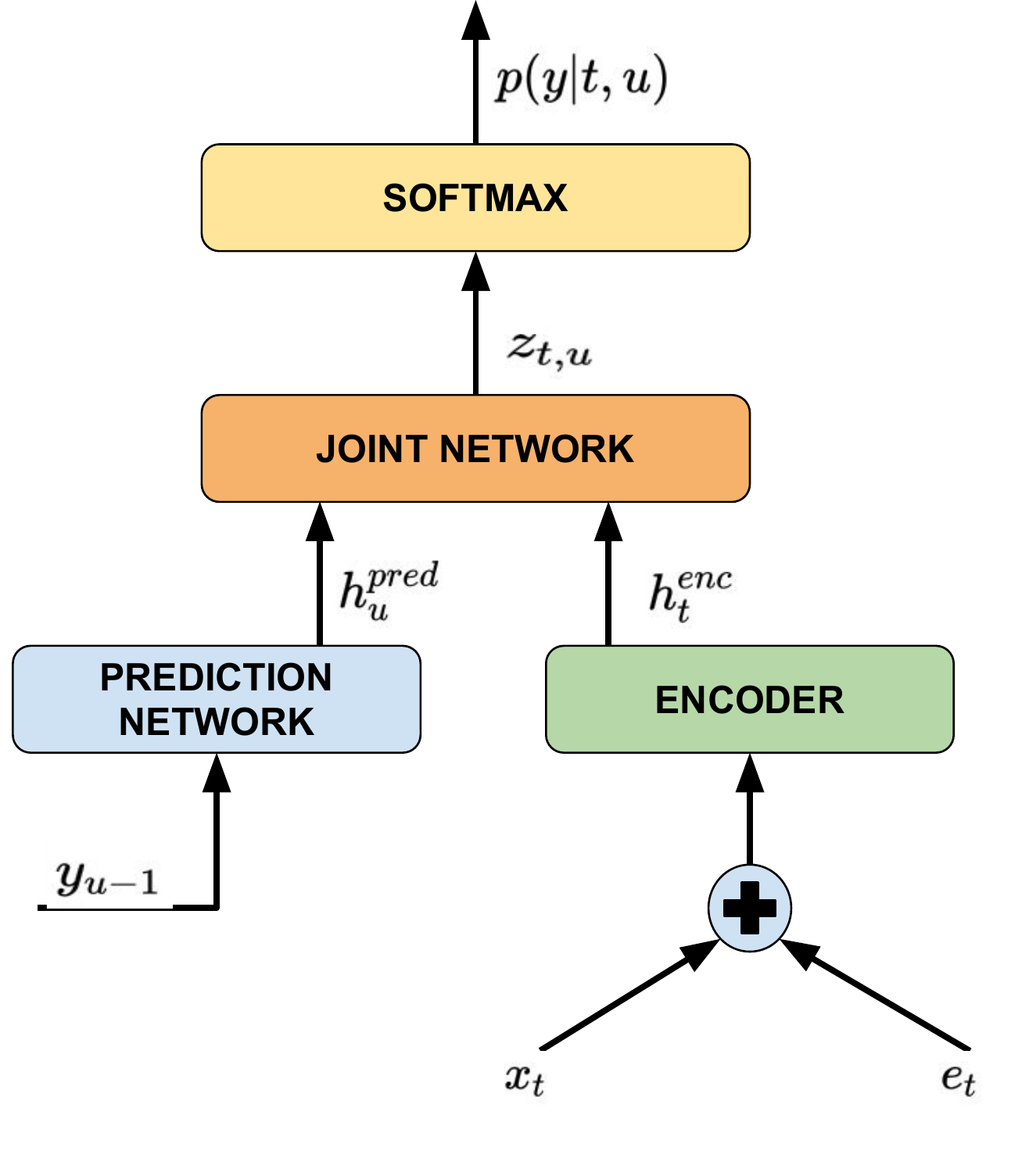}
  \caption{\small{\textit{Incorporating context embeddings into RNN-T ASR. Audio features at each frame are concatenated with $e_t$ which is either per-frame context embeddings or one-hot vector}}}
  \label{fig:incorporate-context-embedding}
\end{figure}

\section{Context Representation with RNN-T}
\label{sec:context-representation}
 In order to use contextual information such as date-time and geo-location in RNN-T ASR, it first needs to be transformed from textual representation to continuous representation. RNN-T model consists of encoder network $h^{enc}$, prediction network $h^{pred}_u$ and joint network $z_{t,u}$.  A typical RNN-T network follows the following operations:

\begin{align}
h_{t}^{enc} = f^{enc}(x_{1}, x_{2}, \cdots, x_{t}) \\
h^{pred}_u = f^{pred}(y^{pred}_1, \cdots, y^{pred}_{u-1}) \\
z_{t,u}=f^{join}(h^{enc}_{t}, h^{pred}_{u}) \\
P(y_{u}|x_1, \cdots, x_t, y_1, \cdots, y_{u-1}) = softmax(z_{t,u})
\end{align}

where $x_1 \cdots x_{t}$ are the audio feature inputs to the RNN-T encoder and $y_1 \cdots y_u$ are the corresponding label sequence.

We extend this by adding contextual information to the encoder network by concatenating feature vector $x$ with context vector $e$. Context vector $e$ can be derived or presented to the network in multiple ways, such as:
\begin{enumerate}
\item {one-hot representation of the context ($o$)}
\item {transforming context ($c$) using a contextual embedding matrix $W$}
\item {feature engineered constant sized vectors ($f$).}
\end{enumerate}

The advantage of representing context as embedding is that it provides the flexibility of combining multiple contextual signals in a lower dimensional space. In our experiments, we use 64 dimensional contextual embeddings. 

\begin{align}
h^{enc}_{ctx_{one-hot}} = f^{enc}(x;o) \\
h^{enc}_{ctx_{embed}} = f^{enc}(x;Wc)\\
h^{enc}_{ctx_{feature-engg}} = f^{enc}(x;f)
\end{align}

\subsection{Date Time as Context}
\label{sec:context-date-time}
A typical date-time information in our dataset looks like - $2020-01-01 T 13:21$. We extract the following information from this datum:
\begin{center}
Hour - 13,
Weekday - Wednesday,
Week No. - 1,
Month - 1
\end{center}
 
In order to bias RNN-T recognition with temporal information, we consider two methods to convert the above information into a continuous vector representation.

\subsubsection{Embedding Representation}
\label{sec:embedding-date-time}
In this method, we learn embedding matrices for hour (24), weekday (7), week number (53) and month (12), where the numbers in the bracket indicate the maximum number of embedding vectors we use for representing corresponding information. These contexts are passed through an embedding layer to generate contextual vectors which are then averaged to represent the complete date-time information. This averaged embedding is used as biasing signal within RNN-T, and is learnt along with RNN-T model training. Assuming embedding vectors of hour, weekday, week number, and month to be  $h_{t}$, $wd_{t}$, $wn_{t}$ and $m_{t}$ respectively, then
\begin{align}
e_{t} = (h_{t} + wd_{t} + wn_{t} + m_{t})/4
\end{align}
In the rest of the paper, this embedding method is addressed as TimeEmbeddingLookUp
\subsubsection{Positional Encoding}
\label{sec:positional-date-time}
 The above embedding approach (Sec  \ref{sec:embedding-date-time}) does not explicitly encode the temporal proximity and cyclical nature of time information. To capture this, we represent the date-time information using an 8-dimensional feature-engineered vector following \cite{martinez2021attention}:

$$
\begin{bmatrix} 
sin(\dfrac{2\pi . hour}{24}), & cos(\dfrac{2\pi . hour}{24}) \\
sin(\dfrac{2\pi . weekday}{7}), & cos(\dfrac{2\pi . weekday}{7}) \\
sin(\dfrac{2\pi . weeknum}{53}), & cos(\dfrac{2\pi . weeknum}{53}) \\
sin(\dfrac{2\pi . month}{12}), & cos(\dfrac{2\pi . month}{12})\\
\end{bmatrix}
\quad
$$
The above representation can clearly express the repetitive behaviour of temporal information. In the following sections, this embedding method is referred to as TimePositionalEncoding.
\subsection{Location as Context}
\label{sec:context-location}
In this work, we used location information up to the state level in the US. The state information for utterances are collected from de-identified user specified information. Given that accent typically varies across the US states, location information is a strong signal to adapt the model to learn these variations. Instead of using all available location information, we clustered utterances with location information to form 20 clusters. The number of clusters are decided empirically with the objective of avoiding multiple centres getting mapped to the same state.
 Approximate geo-location information available from latlong\footnote{https://www.latlong.net/category/states-236-14.html} is used to obtain the state-level geo-location, and euclidean distance is used as the distance metric to learn the cluster centroids. With this we got 20 cluster centroids which are closer in distance. The clusters also include locations outside the US, which correspond to small percentage of users using the devices outside the main region. For some utterances, the location information is not available and we assign it to None cluster. We explored transforming geo-location using an embedding layer (GeoEmbeddingLookUp), and also encoding it as a one-hot vector (GeoOneHot) to bias the RNN-T model.

\subsection{Combination of Context}
\label{sec:context-combined}
We also ran experiments combining date-time and location information to bias the RNN-T search. We expect the date-time and location together to be a much stronger signal than either individual signals alone. We use embedding approach to combine the context (CombinedTimeGeo), where the  embedding matrix is learned to map combined context into lower dimension contextual embedding vector.

\section{Data and Experimental Setup}
\label{sec:setup}
\subsection{Datasets}

For our experiments, we used de-identified human-labelled speech data collected from queries to voice controlled far-field devices. The dataset was randomly split into train, dev and eval. The training set comprised of 30K hours of de-identified human-labelled US English recordings. Each recording includes meta information such as time stamp and optional US state from which it originated. The eval set consists of approximately 100 hours of generic utterances.  We also evaluate our models on a communication specific test set of 23 hours of utterances. Both the evaluation test sets are mutually exclusive. We refer to the former as Eval test set and the latter as Comms test set.

\subsection{Experimental Setup}
\label{sec:model-details}

\subsubsection{Full Resource RNN-T ASR}

The baseline RNN-T model consists of 5 encoder layers of 1024 hidden units, with a final layer output dimension of 512. The prediction network has an embedding layer of 512 units, 2 LSTM layers of 1024 units, and a final output dimension 512. The joint network is a feed forward network of 512  hidden units and a final output dimension of 4001. The 4001 dimensional output, corresponds to the number of subword tokens, is passed through a final softmax layer. The subword vocabulary was generated using the byte pair encoding algorithm \cite{shibata1999byte}

The contextual RNN-T ASR model has an additional embedding layer generating embedded representation of 64 dimensions which are appended to the input of the encoder at every time step. The two exceptions being:
\begin{enumerate}
\item{the positional time encoding has an 8 dimensional context vector}
\item{one-hot geographical information has a 21 dimensional context vector}
\end{enumerate}

All the models are trained on 30K hours of training data. 

\subsubsection{Low Resource RNN-T ASR}

We also trained both the baseline and contextual models on 10K hours of data. The main motivation for this study is to analyze the effect of context in the low training data regime. In order to prevent over-fitting, we scaled down the number of parameters of the models. Number of hidden units of both the encoder and decoder layers were reduced to 760. The feed-forward joint network is also removed. The encoder and decoder outputs are summed and provided as an input to the softmax layer. All other specifications are kept consistent with the full-resource models.

Both full-resource and low-resource models use a 64-dimensional log filter bank energy features computed over 25ms window with 10ms shift. Each feature vector is stacked with 2 frames to the left and down sampled to a 30ms frame rate. We also augment the acoustic training data with SpecAugment \cite{park2019specaugment} to improve the robustness. All models are trained using the Adam optimizer \cite{kingma2014adam}, with a learning rate schedule including an initial linear warm-up phase, a constant phase, and an exponential decay phase \cite{chen2018best}. These hyper-parameters are not specifically tuned for this work.

\section{Results and Discussion}
\label{sec:results-and-discussion}

\subsection{Overall WER Comparison}

Table \ref{tab:WERR-breakdown} shows the overall Relative Word Error Rate Reduction (WERR) with respect to the baseline RNN-T model without context. Performance of our baseline system is below 10\% WER absolute. The magnitude of improvement on Comms test set is more than that of Eval test set, which signifies that these contextual signals are more favourable for communication specific utterances. Overall, incorporating geo-location information as one-hot provides the maximum WERR of 3.48\%. Based on the performance, we chose TimeEmbeddingLookUp and GeoOneHot models for further analyses. 

\begin{table}[h]
	\centering
	\small
	\caption{\small \textit{{Relative WERRs of full resource contextual models w.r.t baseline}}}
	\scalebox{1.0}{
		\begin{tabular}{c|c|c|c }
			\toprule
			Model & Eval & Comms & \#params      \\
		     \midrule
			Baseline & --- & --- & 58.4M \\
			\midrule
			\textbf{TimeEmbeddingLookUp} & \textbf{1.73\%} & \textbf{2.68\%} &58.7M  \\
			TimePositionalEncoding& 1.33\%& 2.39\% & 58.4M  \\
			\midrule
			GeoEmbeddingLookUp & 1.47\% & 1.6\% & 58.6M  \\
			\textbf{GeoOneHot} & \textbf{2.27\%} &  \textbf{3.48\%} &  58.5M \\
			\bottomrule
	\end{tabular}}
	\label{tab:WERR-breakdown}
\end{table}

\subsection{Domain-wise WER Comparison}
\label{sec:domain-wer}
We show the WER improvement of Eval set for various domains in Table \ref{tab:per-domain-WERR-breakdown}. In general, we see gains on all top domains of interest with both approaches. Geo-location exhibits superior performance in domains like Music and CallingAndMessaging etc. These domains capture region specific preferences for music and video along with accent variations of proper nouns. On the other hand temporal context shows improvement on domains where queries come mostly at a certain point of time in a day, e.g. DailyBriefing, Weather etc. 

\begin{table}[h!]
\centering
	       \small
	       \caption{\small \textit{Per-domain relative WERR (\%) breakdown on top 10 frequent domains selected from the test set. Analysis was done using the full resource model.}}
	       \scalebox{1.0}{
		               \begin{tabular}{l c  c }
			                       \toprule
			                       Domain                                   &  TimeEmbeddingLookUp                  & GeoOneHot     \\
			                       \midrule
			                        \textbf{Music}                                  & 1.72          & \textbf{3.57} \\
			                       Shopping                           & 2.03        & 1.49  \\
			                       \textbf{CallingAndMessaging} &  2.04       & \textbf{5.68} \\
			                       Global                           & 0.86          & 2.58 \\
			                       \textbf{DailyBriefing} & \textbf{4.55} & 0.91 \\
			                       Knowledge                        & 2.14                  & 1.97 \\
			                       \textbf{Video}     & 3.13  &       \textbf{6.26} \\
			                       \textbf{Weather}     & \textbf{5.92}  & 4.67 \\
			                       Information                       & 1.44                  & 1.96 \\
			                       ScienceAndTechnology	& 	 -0.49 &	-2.17 \\
			                       \bottomrule
			               \end{tabular}}
	       \label{tab:per-domain-WERR-breakdown}
	\end{table}

\subsection{Combined Context}
\label{sec:combined-context}
\begin{table}[!h]
	\centering
	\small
	\caption{\small \textit{{Relative WERRs of combined contextual model w.r.t baseline. Both the models are trained on complete 30K hours of data}}}
	\scalebox{1.0}{
		\begin{tabular}{c | c c}
			\toprule
			Model & Eval & Comms      \\
			 \midrule
			Baseline & --- & --- \\
			\midrule
			\textbf{CombinedTimeGeo} & \textbf{3.6\%} & \textbf{4.62\%} \\
			\bottomrule
	\end{tabular}}
	\label{tab:Combined-WERR-breakdown}
\end{table}

The effect of combining the two contextual signals is captured in Table \ref{tab:Combined-WERR-breakdown}. Combining the contextual information shows superior performance compared to individual contextual models (Table \ref{tab:WERR-breakdown}). This establishes the additive effect of the location and date-time signals. In domain-wise study, we see a similar additive effect on several domains like CallingAndMessaging ($6.05\%$), Knowledge ($4.44\%$),  Video ($7.86\%$) and Weather ($11.56\%$) etc. Moreover, in domains like ScienceAndTechnology, where neither of the individual contextual models showed any improvement, the combined model performed significantly better ($6.61\%$ WERR).

\subsection{Low Resource Simulation}
In practice we often face data scarcity while developing ASR for a new language or locale. In such cases, we can easily leverage contextual information as they are readily available to gain additional performance benefits. We simulated this situation by randomly selecting a 10K hours subset from the full training data, and trained both the baseline and the contextual models on this subset. The model sizes have also been reduced to avoid over-fitting as described in Section \ref{sec:model-details}. Table \ref{tab:LowResource-WERR-breakdown} shows that the magnitude of gains have increased as compared to full resource, which demonstrates the efficacy of these contextual signals for low resource scenarios.

\begin{table}[t]
	\centering
	\small
	\caption{\small \textit{{Relative WERRs of Low Resource contextual models wrt baseline}}}
	\scalebox{1.0}{
		\begin{tabular}{c|c|c|c }
			\toprule
			Model & Eval test set    & Comms test set & \#params      \\
			\midrule
			Baseline & --- & --- &  38M\\
			TimeEmbeddingLookUp& 4.03\%& 3.66\% &  38.2M \\
			GeoOneHot & 4.29\% & 3.76\% &   38.1M\\
			\bottomrule
	\end{tabular}}
	\label{tab:LowResource-WERR-breakdown}
\end{table}

\subsection{Month and State-wise WER Comparison}
\label{sec:month-state-wer}
 \begin{table}[t]
	\centering
	\small
	\caption{\small \textit{{WERR (\%) for top 3 and bottom 3 performing month and geographical state/country}}}
	\scalebox{1.0}{
		\begin{tabular}{c|c|c }
			\toprule
			Model & Top 3 (WERR)   & Bottom 3 (WERR)      \\
			\midrule
			& December (11.61) & November (0.25)\\
			TimeEmbeddingLookUp & February (10.49)& August (-0.64)\\
			& January (5.24) & September (-4.86)\\
			\midrule
			& Germany (9.48) & Ohio (1.52)\\
			GeoOneHot & Hawaii (5.38) & Florida (1.02)\\
			& Washington (4.77) & California (0.79)\\
			\bottomrule
	\end{tabular}}
	\label{tab:top-bottom-3-performance}
\end{table}

To further understand the effect of our proposed methods, we performed a month-wise and state-wise WERR analyses for date-time and geo-location based models respectively. We have captured the best and worst performing month/state for date-time/geo-location models in Table \ref{tab:top-bottom-3-performance}. The date-time information enhances the performance of RNN-T for some winter months considerably. On the other hand, geo-location signal significantly enhance the performance on utterances coming from low resource regions like Germany, Hawaii and its nearby locations. This can be mostly attributed to difference in acoustics of utterances coming from these regions as they are different from that of other regions captured by the geo-location clusters. Even for the worst performing region, we do not see any degradation of performance with geo-location as context.

 \begin{table}[ht!]
	\centering
	\small
	\caption{\small \textit{{Comparison of contextual model output and baseline output}}}
	\scalebox{1.0}{
		\centering
		\begin{tabular}{P{0.25\textwidth}| P{0.25\textwidth} | P{0.25\textwidth} | P{0.2\textwidth}}
			\toprule
			Reference  &  Baseline output   &  Contextual output & Context used \\
			\midrule
			good night and \textbf{happy hanukkah} & good night and happy hobbiter & good night and \textbf{happy hanukkah}  & Time \\
			\midrule
			what's the \textbf{christmas cat story} & what's the christmas car story & what's the \textbf{christmas cat story} & Time \\
			\midrule
			call \textbf{guillermo} & call galermo & call \textbf{guillermo}  & Geo-location \\
			\midrule
			turn to my \textbf{kirk franklin} radio & turn to my park franklin radio & turn to my \textbf{kirk franklin} radio & Geo-location \\
			\bottomrule
	\end{tabular}}
	\label{tab:context-output-examples}
\end{table}

\subsection{Baseline and Contextual Model Outputs}

In Table \ref{tab:context-output-examples} we compare a few example predictions from baseline and contextual model. We see that, the time information helps in recognition of phrases like ``hanukkah" and ``christmas cat", which the baseline model fails to recognize. These phrases are seen in December which is implicitly captured by the contextual model.

Similar to time, when we use location information as context, it captures the accent variations and local preferences of music and videos. The contextual model was able to capture the local accent variation and correctly output ``guillermo" while the baseline model outputs ``galermo" which is somewhat phonetically similar to the correct phrase. Similarly ``kirk franklin" was correctly recognized compared to incorrect baseline output of ``park franklin" which shows that the model was able to capture local variation in music preferences without any external supervision.

\begin{figure}[hb!]
\centering
  \includegraphics[scale=0.5,width=0.7\linewidth]{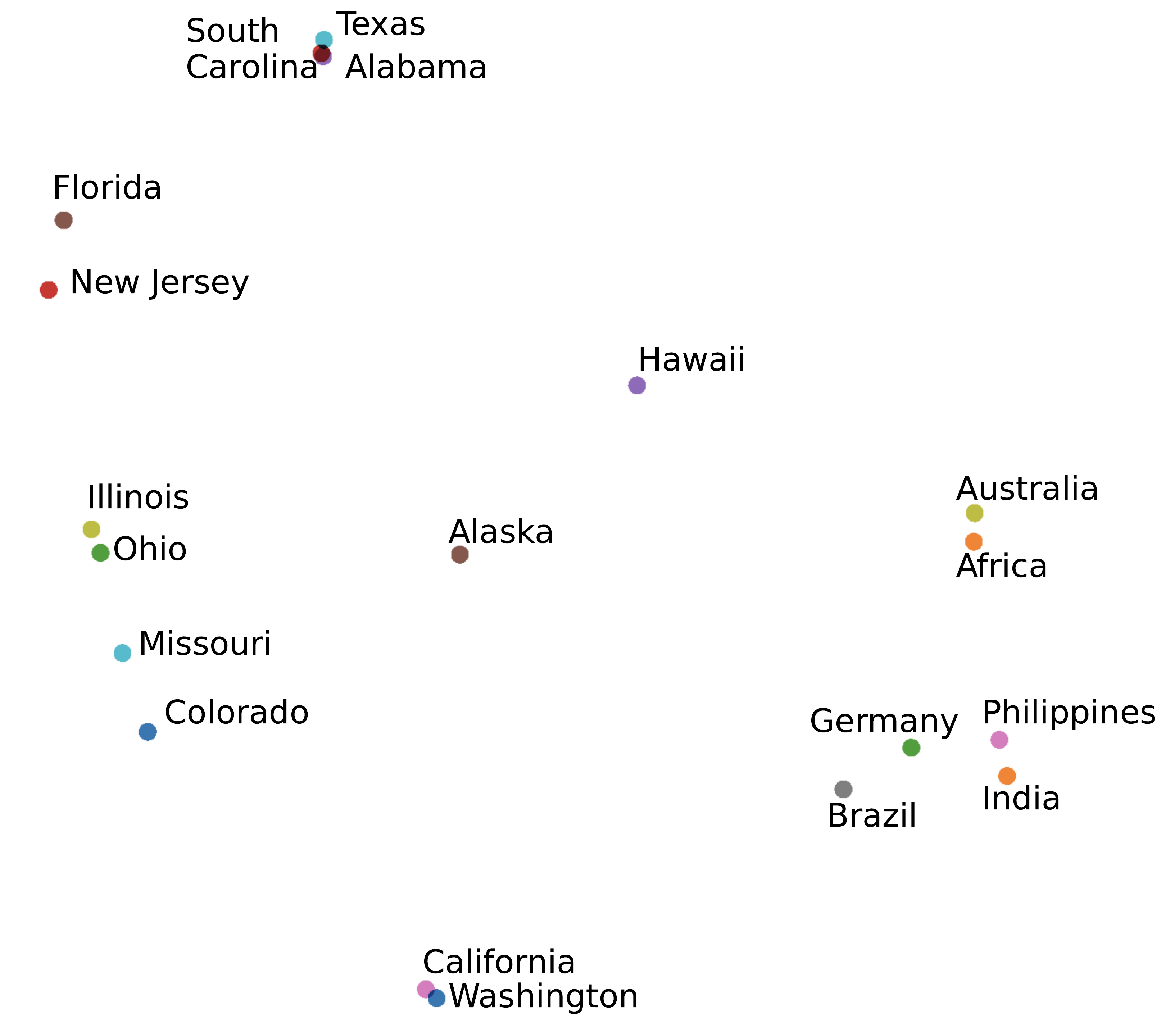}
  \caption{\small{\textit{t-SNE plot for geo-embeddings}}}
  \label{fig:tsnePlot-geo}
\end{figure}

\subsection{t-SNE Plot Analysis}
\label{sec:tsne-plot-analysis}

Embeddings are meant to capture some implicit information about the context it represents. To understand the significance of the embedding vectors learnt by the models, we projected the 64 dimensional geo-location and month embedding vectors from GeoEmbeddingLookUp and TimeEmbeddingLookUp on 2-D space using t-SNE with default parameters of Embedding Projector\footnote{https://projector.tensorflow.org/}.

In Figure \ref{fig:tsnePlot-geo} we show the t-SNE plot for the learnt geo-embedding vectors. We can see that the geographically close states in the US have formed clusters in the embedding space (e.g - California:Washington, Illinois:Ohio etc.) which seems to capture local variations like regional movies and song preference and also local accents. We can also see a clear demarcation between the US locations and the non-US locations like Brazil, Germany and India. Users across the US and the non-US will have different accents which are captured by the model. This proves that the geographical location distribution is important for the model and the model has learnt that without any external supervision.

\begin{figure}[t]
	\centering
	\includegraphics[scale=0.5,width=0.7\linewidth]{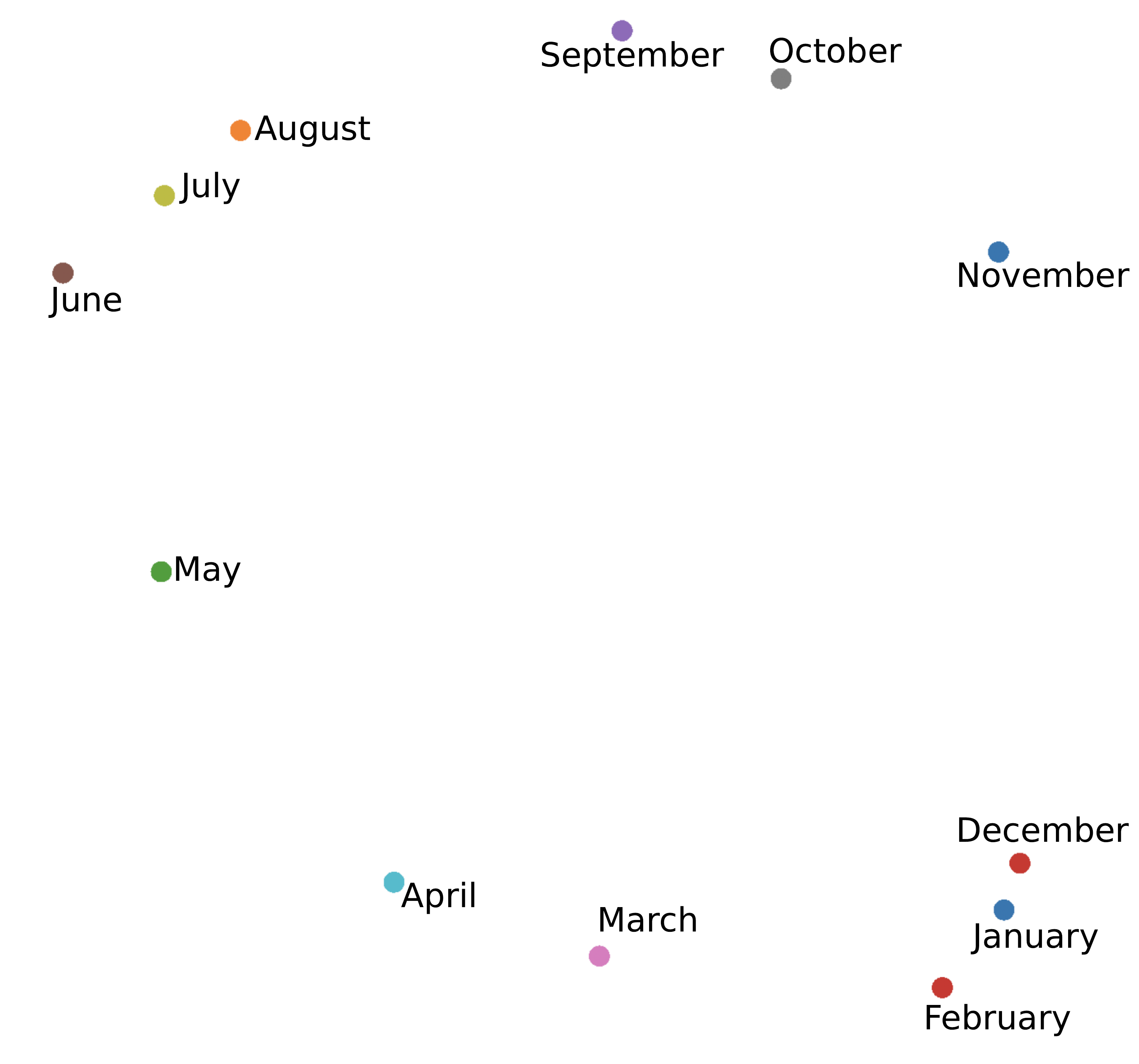}
	\caption{\small{\textit{t-SNE plot for month embeddings}}}
	\label{fig:tsnePlot-month}
\end{figure}

Figure \ref{fig:tsnePlot-month} shows the t-SNE plot for embedding vectors corresponding to month. We can observe a clear temporal ordering among the learnt month vectors which demonstrate the capacity of our models to implicitly learn the ordering among months from data. This phenomenon also proves that the temporal ordering of months is somewhat important for the task of ASR. Note that, we have not imposed any ordering constraint on any of our models. 

\section{Conclusions}
\label{sec:conclusion}
In this paper, we explored the benefits of using contextual signals to improve the overall performance of end-to-end ASR based on RNN-T. We demonstrated the effectiveness of date-time and location as context by building  ASR models on 30K and 10K hours of data. We provided empirical evidence that biasing ASR using contextual signals improves the overall accuracy. The use of individual contextual signals improved the ASR WER up to 3.48\% relative, and where as their combination resulted in about 4.62\% relative gain. Our analysis with t-SNE plot of embedding vectors for both geo-location and date-time context showed that the model was able to extract meaningful information from these signals and improving ASR, thereby possibly reducing the need for additional training data which is now critical for performance improvement. As a part of future work, we would like to add dynamic contextual signals along with these static ones to further enhance RNN-T ASR performance.

\bibliographystyle{splncs04}
\bibliography{paper}

\end{document}